# Beyond Fidelity: Evaluating Synthetic Medical Imaging Data

Seyed Kahaki*, Alexander R. Webber, Tahsin Rahman, Elim Thompson, Jana Delfino, Aldo Badano

Division of Imaging, Diagnostics, and Software Reliability, U.S. Food and Drug Administration (FDA)

## ABSTRACT

The growing use of synthetic medical imaging datasets (SMD) in the development and evaluation of machine learning and deep learning models has motivated interest in systematic methods to assess dataset quality across a variety of imaging modalities. Common evaluation strategies rely heavily on fidelity-oriented metrics developed for general computer vision and may not capture clinically relevant properties such as anatomical plausibility, dataset coverage, metadata completeness, subgroup consistency, or documentation quality. We present the ScoreCard Toolkit, an open-source method that operationalizes six of the seven dimensions found within the 7Cs approach for SMD evaluation across Congruence, Coverage, Completeness, Consistency, Constraint, and Comprehension. For this work, Compliance is considered out of scope. The ScoreCard Toolkit provides a modular pipeline for preprocessing, feature extraction, metric computation, visualization, and structured scorecard generation. This approach complements standard fidelity-based evaluation by exposing quality characteristics not captured by those measures alone, including constraint violations, anatomically implausible shape artifacts, and metadata-related consistency issues. The toolkit supports appearance, texture, and shape features and is designed to accommodate additional feature representations, including radiomic features, deep-learning embeddings, and foundation-model representations. We demonstrate the end-to-end workflow across three imaging modalities: synthetic digital pathology whole-slide image (WSI) patch data, synthetic digital mammography, and synthetic head computer tomography (CT). Across all three modalities, the resulting scorecards reveal dataset-specific quality characteristics of the SMDs, such as distributional biases, lack of variety, and coverage gaps that may impact their effectiveness when used for model development or testing. ScoreCard provides a practical, extensible, and modality-agnostic approach for evaluating and transparently reporting SMD quality in medical imaging applications. The code is freely available at https://github.com/DIDSR/ScoreCard.

*Corresponding Author: seyed.kahaki@fda.hhs.gov

## INTRODUCTION

The limited availability of high-quality annotated medical imaging data remains a challenge for the development and evaluation of computational-driven diagnostic tools [1]. As defined in the educational resource FDA Digital Health and Artificial Intelligence Glossary, synthetic medical imaging datasets (SMD) are artificially generated data which are intended to reproduce relevant statistical and clinical characteristics of patient data [2]. SMD may help address data scarcity, particularly for rare diseases or underrepresented populations [3]. However, the usefulness of synthetic data depends not only on visual realism, but also on whether the generated data preserve characteristics relevant to the intended application. Consequently, evaluation of SMD quality has been identified as a relevant consideration before synthetic data are used for model development or assessment.

Existing synthetic-image evaluation frequently emphasizes statistical fidelity or perceptual similarity. Metrics such as Fréchet Inception Distance (FID) and Inception Score (IS) can provide useful measures of distributional or image-level characteristics, but they do not directly evaluate many properties that may be important in medical imaging, including clinical constraints, anatomical plausibility, demographic or subgroup consistency, metadata completeness, and documentation of the generation process [4, 5]. For example, a synthetic image may appear realistic or achieve favorable fidelity metrics while containing an anatomically implausible structure or exhibiting a systematic bias in anatomical shape that could affect downstream model results [6]. These observations motivate evaluation strategies that characterize multiple, complementary dimensions of synthetic data quality rather than relying on a single fidelity measure.

To address this gap and support the growing scientific need for a transparent evaluation approach, we applied the 7Cs approach described in Zamzmi et al. [7, 8], and developed a toolkit grounded in synthetic data evaluation. As synthetic data is increasingly explored for use in the development and evaluation of data-driven computational tools in medical imaging, systematic methods to assess data quality and reliability are an active area of investigation. ScoreCard extends beyond fidelity to encompass seven clinically relevant quality dimensions: Congruence, Coverage, Completeness, Consistency, Constraint, Compliance (defined here as dataset providers' responsibility for applicable regulations and institutional policies, and not quantified by the toolkit), and Comprehension. By providing a holistic view of data quality, the toolkit provides a quantitative characterization of synthetic dataset properties that may inform, but does not substitute for, expert judgment regarding dataset suitability.

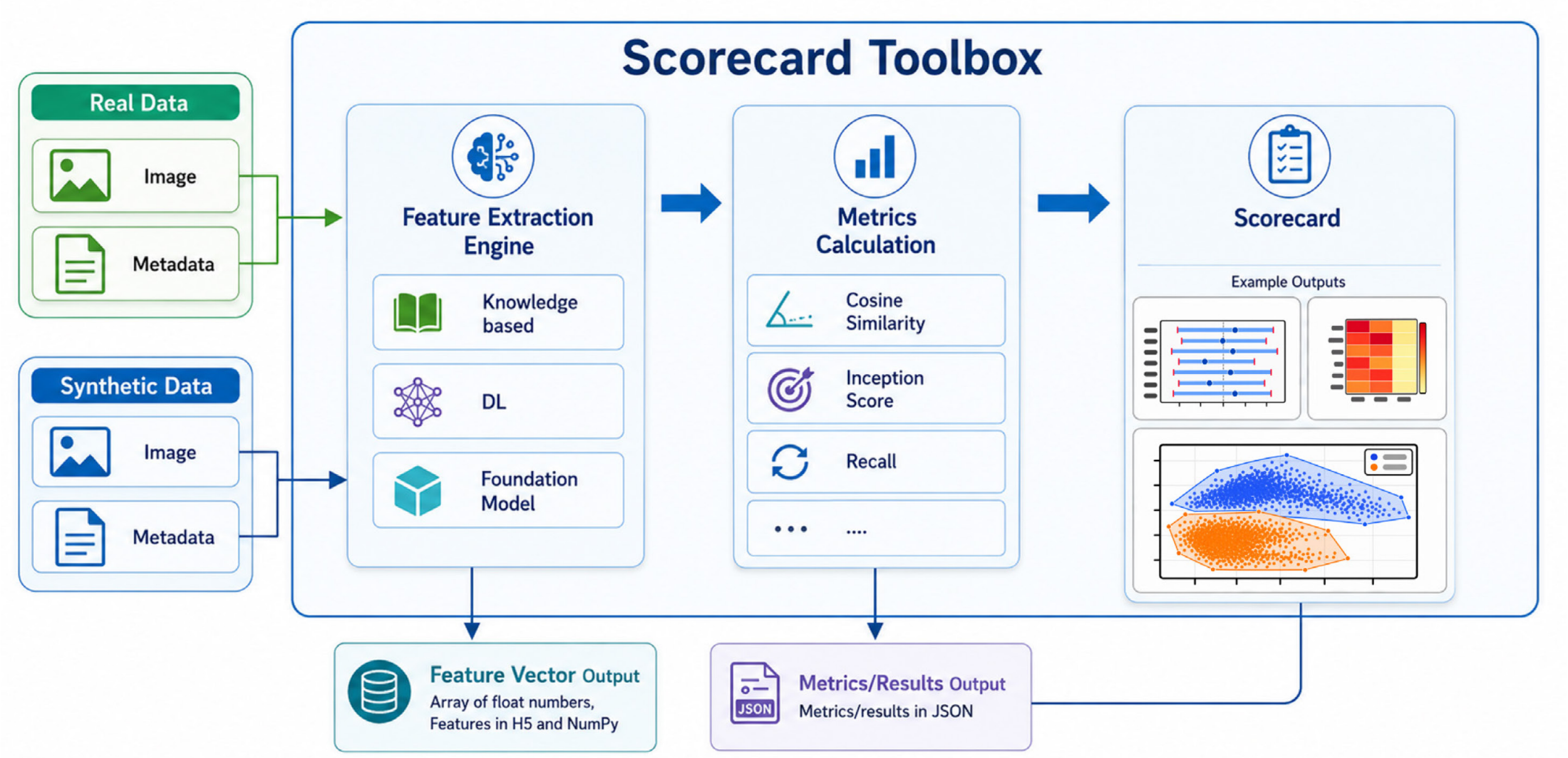

Figure 1: **Overview of the Synthetic Data ScoreCard Pipeline**. The pipeline accepts patient and synthetic data (including PNG images, NIfTI volumes, shape masks in .mat/HDF5 format, and metadata in .csv/JSON/XML format) as inputs. A Feature Extraction Engine processes both patient and synthetic data, producing intermediate feature representations. These features are passed to a Metrics Calculation module that computes scores across the various synthetic data evaluation quality dimensions. Results are stored in JSON format and used to generate the final Scorecard report. Feature extraction strategies may include knowledge-based methods (texture, color, GLCM, Gabor), deep learning models (ResNet-50), and general-purpose foundation models (DINOv2).

As illustrated in Figure 1, the toolkit implements an automated, modular pipeline for feature extraction, metric computation, and scorecard report generation. To demonstrate its utility and generalizability across imaging modalities, we applied the toolkit to three distinct datasets: synthetic digital pathology whole-slide image (WSI) patch data, synthetic digital mammography, and synthetic computer tomography (CT). These modalities differ substantially in image dimensionality, acquisition physics, anatomical content, and the types of quality failures that generative models are known to produce, making them a meaningful test of the toolkit's portability. Ultimately, the goal of this work is to provide a practical, open-source tool that enables SMD developers, researchers, and stakeholders to systematically assess and transparently report synthetic data quality across a wide range of medical imaging modalities.

## METHODS

### ScoreCard Toolkit Architecture

The ScoreCard Toolkit is organized as a modular pipeline with three core components: (1) a Feature Extraction Engine, (2) a Metrics Calculation Module, and (3) a ScoreCard Report Generator. The pipeline accepts patient and synthetic image data together with available metadata. It produces intermediate feature representations stored in HDF5 formats, metric results in JSON or CSV format, and a structured scorecard report in either PNG or SVG. This separation of feature extraction, metric calculation, and reporting is intended to allow modality- or application-specific components to be added without resulting in changes to the complete evaluation workflow. Figure 1 illustrates the pipeline applied across the three imaging modalities demonstrated in this work.

Before feature extraction, patient and synthetic datasets undergo the same preprocessing steps to support consistent comparison. The modular pipeline permits dataset- or modality-specific preprocessing operations to be incorporated when required. For WSI data, images are normalized across color channels to reduce staining variation and are tiled into non-overlapping patches [9]. For 2D digital mammography data, preprocessing follows the procedures specified by the respective dataset authors. Images are then resized and intensity-scaled to ensure uniformity across acquisitions. For 3D head CT data, the raw CT pixel values are clipped to a range of -1024 to 1024 Hounsfield Units (HU) and linearly normalized to 8-bit integers (0–255) for processing. Features were extracted from the entire 3D volume for every single case, with feature analysis focused on the location in the volume delineated by a mask of the blood arteries.

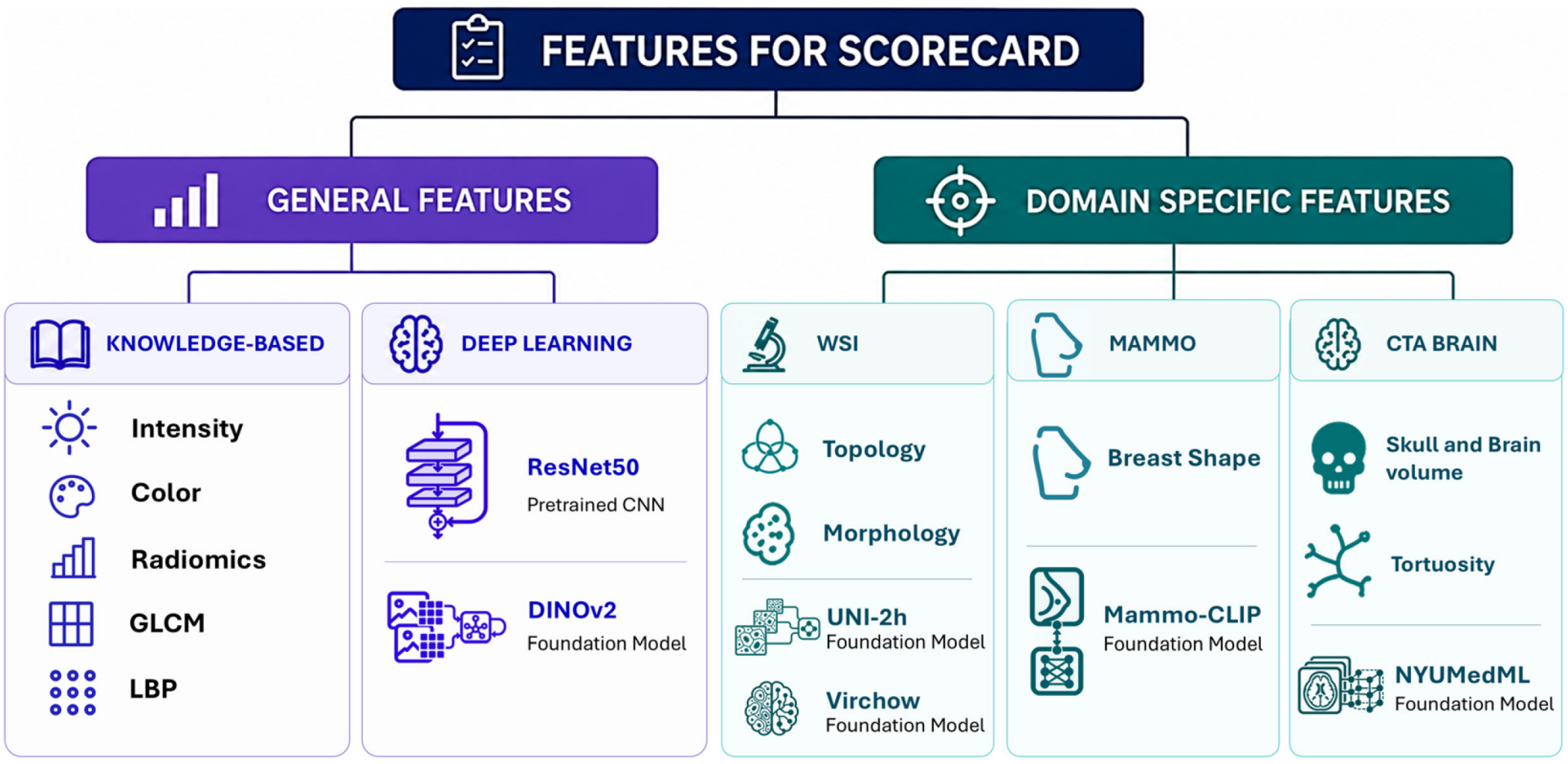

Figure 2: **Overview of the features that can be utilized in this pipeline**. General features that can be utilized across data modalities include knowledge-based descriptors (intensity, color, radiomics, GLCM, LBP) and deep learning models (ResNet50, DINOv2). Domain-specific features tailored to whole slide imaging (WSI), mammography (MAMMO), and head CT modalities, are also available, incorporating both handcrafted and foundation model-derived representations.

## Feature Extraction

Following preprocessing, data is passed to the Feature Extraction Engine. The ScoreCard toolkit can be used with a wide variety of feature types, shown in Figure 2, starting from domain-agnostic knowledge-based and deep learning features and moving to domain-specific features such as cell topology, breast shape, and blood vessel tortuosity. The Feature Extraction Engine is used to generate feature vectors of those features specified by the user for each sample of the preprocessed data, which are then stored in an HDF5 file format for analysis and metric calculation.

For WSI data, the current implementation includes low-level, knowledge-based descriptors such as Gray Level Co-occurrence Matrix (GLCM) texture features (contrast, homogeneity, energy, and correlation), mean RGB, and mean intensity features. When segmentation masks are available, nuclei-level shape descriptors can also be evaluated, including convexity, roundness, solidity, eccentricity, and skeleton features, as well as topological features such as nearest neighbor based on Voronoi and radial distance neighborhoods. Additionally, features can be extracted using foundation models, such as UNI-2h [11].

For digital mammography data, in addition to the knowledge-based features used in WSI, a DINOv2 [10] foundation model feature extractor is available to extract deep learning features from the images. DINOv2 is a general-purpose visual foundation model developed by researchers at Meta AI that produces image features transferable across a wide range of downstream tasks without task-specific fine-tuning. It is built on the Vision Transformer (ViT) architecture and was pre-trained on LVD-142M [10], a curated dataset of 142 million images using a self-supervised teacher-student distillation approach. In this work, DINOv2 was applied as a feature extractor for digital mammography data; its applicability across other radiology domains was not evaluated in this study.

Finally, for head CT data, both mean intensity and the same GLCM texture feature family as those used in the WSI and mammography pipelines (contrast, homogeneity, energy, and correlation) are available. For all modalities, Principal Component Analysis (PCA) is applied to the feature vector to support visualization and distributional comparison across acquisition conditions.

Selecting features matched to the imaging modality, available annotations, and quality characteristics being assessed can improve the relevance of the evaluation. The pipeline is designed to accommodate additional feature classes, including statistical features, radiomic features, deep-learning embeddings, and foundation-model representations such as DINOv2 and UNI-2h. A list of all features available for each modality in the ScoreCard tool kit can be found in Supplement 1.

## Synthetic Data Evaluation and Metrics

ScoreCard uses the extracted representations to calculate quantitative measures of synthetic data quality relative to patient reference data for the applicable dimensions of the 7Cs approach [7]. Congruence characterizes alignment between synthetic and patient feature distributions and is assessed using measures including Cosine Similarity, Earth Mover's Distance, and Jensen-Shannon Divergence. Coverage characterizes the variability and range represented by the synthetic data relative to patient data using measures including Convex Hull Volume, Variance, Entropy, and Distance to Centroid. These measures are intended to provide complementary views of distributional alignment and representation rather than a single universal quality score.

Constraint measures adherence to known clinical, anatomical, or geometric constraints. To quantify this, we formalize the Constraint Violation Rate (CVR). Let $S = \{s_1, s_2, \ldots, s_N\}$ represent the set of $N$ synthetic samples, and $C = \{c_1, c_2, \ldots, c_K\}$ represent a set of $K$ modality-specific clinical constraints. A binary indicator function $\mathrm{I}(s_i, c_k)$ sets to 1 if sample $s_i$ violates constraint $c_k$, and 0 otherwise. The overall CVR is computed as the proportion of synthetic samples that violate at least one constraint: $\mathrm{CVR} = \frac{1}{N}\sum_{i=1}^{N} \max_k \mathrm{I}(s_i, c_k)$. For WSI data, constraints include cell morphology rules such as the absence of non-physiological cell shapes. For other modalities, such as head CT data, constraints may include skull geometry thresholds and lesion size and shape plausibility. Completeness assesses whether the synthetic dataset contains all information necessary for its intended application, computed as the Proportion of Expected Fields and Missing Data Percentage relative to a patient reference dataset. Consistency evaluates the uniformity of SMD Completeness across demographic subgroups or over time, measured using Variance, Maximum-Minimum Difference, and Analysis of Variance (ANOVA).

Comprehension is assessed qualitatively using a set of criteria that captures the transparency and clarity of the data-generation process and the quality of accompanying documentation. Compliance is not quantified by the current toolkit. Instead, dataset providers remain responsible for determining and documenting compliance with applicable regulations, institutional policies, and ethical responsibilities. ScoreCard therefore reports Compliance as an evaluation consideration rather than inferring regulatory or policy compliance from image-derived metrics.

## Scorecard Report

After feature extraction and metric computation, the toolkit generates a structured report for each evaluated synthetic dataset. The report includes dataset-level information, example images, feature-distribution visualizations, dimension-specific quantitative results where applicable, and qualitative information for Comprehension. The purpose of the report is to retain the individual dimensions and metrics rather than collapse heterogeneous quality characteristics into a single value. This structure allows users to identify which aspects of a synthetic dataset align with the patient reference dataset, and which aspects may warrant additional investigation.

## Implementation

The ScoreCard Toolkit is implemented entirely in Python, ensuring cross-platform compatibility and ease of integration into existing MLOps pipelines. Appearance and texture feature extraction is performed using the PyFeats library [12]. The pipeline includes source scripts to programmatically perform the assessments and save the features and results, as well as Jupyter notebooks for each component to help users familiarize themselves with the different dimensions. A key usability feature of the pipeline is the inclusion of a Python Flask-based application which allows users to access all the features of the toolkit using a simple Graphical User Interface (GUI) in a browser. Users can specify patient and synthetic data directories for feature extraction, visually analyze distributions of the extracted features, and compute all metrics for the ScoreCard report.

Table 1. Description of metrics implemented within the ScoreCard Toolkit

| Component | Description | Example Metrics / Methods |
|---|---|---|
| **Congruence** | Measures the degree to which the distribution of synthetic data aligns with the distribution of provided real patient data. | Cosine Similarity, Jensen-Shannon Divergence, Earth Movers Distance |
| **Coverage** | Evaluates the extent to which SMD captures the coverage and range inherent in provided patient data. | Convex Hull Volume, Variance, Distance to Centroid |
| **Completeness** | Evaluates whether the generated data contains all necessary details relevant to the intended task. | Proportion of Required Fields, Missing Data Percentage (compared to a reference dictionary) |
| **Consistency** | Assesses the stability of SMD Completeness across demographic subgroups, disease classes, or over time. | Variance, ANOVA, Maximum-Minimum Difference |
| **Constraint** | Assesses adherence to known anatomical, biological, geometric, or user-defined constraints. | Constraint Violation Rate (measures how often synthetic features fall outside ranges defined by corresponding real features) |
| **Comprehension** | A set of criteria to evaluate the transparency and clarity of the data generation process and the quality of accompanying documentation. | Qualitative set of criteria (no quantitative metric) |
| **Compliance** | Dataset providers are encouraged to ensure compliance with applicable regulations and institutional policies. | Not evaluated quantitatively by this toolbox; compliance is the responsibility of dataset providers. |

## Dataset Description

### WSI Patches

The ScoreCard Toolkit was tested on a synthetic WSI dataset using the MoNuSeg patient dataset as the patient reference dataset [13]. MoNuSeg contains hematoxylin and eosin (H&E)-stained tissue images at 40x magnification sourced from The Cancer Genome Atlas (TCGA) archive [14]. The dataset includes annotated images from multiple patients and organ tumor types. For the present demonstration, 16 MoNuSeg images of 1000 × 1000 pixels, containing approximately 10,000 nuclei in total, were used as the patient reference data. Synthetic images were generated for this study using HistoGen [15], a histopathology image generative model conditioned on nuclei segmentation masks from the patient image patches. Ten synthetic images were generated per patient image patch, yielding 160 synthetic images of 1000 × 1000 pixels for evaluation. No train/test split was applied because the objective of this experiment was to demonstrate evaluation of synthetic data quality rather than to train or independently test a predictive model.

### Digital Mammography

Three digital mammography datasets were used in this study: one real-world patient dataset and two synthetic datasets.

The patient dataset used was the publicly available EMory BrEast imaging Dataset (EMBED) [16]. EMBED is a large-scale mammography dataset developed and managed by the Health Innovation and Translational Informatics Lab at Emory University, with the goal of addressing persistent gaps in variation, scale, and annotation granularity that have limited the generalizability of breast imaging machine learning models. Collected between 2013 and 2020, EMBED encompasses approximately 3.4 million screening and diagnostic mammographic images from around 116,000 women, and includes roughly 40,000–60,000 annotated lesions linked to structured imaging descriptors and pathologically confirmed outcomes. The dataset includes 2D full-field digital mammograms, synthetic 2D C-view images derived from digital breast tomosynthesis (DBT), and 3D DBT volumes, with additional modalities such as ultrasound and MRI planned for future releases. Approximately 20% of the full dataset is freely available to the research community through the Amazon Web Services (AWS) Open Data Program (mention of commercial products does not constitute FDA endorsement; see Disclaimer). In this work, 3,775 DICOM samples from the publicly available subset of EMBED were used.

The first synthetic dataset, CSAW-syn, consists of 100,000 mammographic images generated using SinKove [17], a generative framework capable of producing synthetic medical imaging data across multiple modalities. For this dataset, a latent diffusion model within the SinKove framework was trained on the CSAW-M mammography patient dataset [18]. The generative architecture is comprised ofa variational autoencoder with Kullback-Leibler (KL) regularization to learn a latent representation from the patient data, followed by a diffusion U-Net to model the latent distributions. Training was performed by the dataset developer using the NVIDIA MONAI framework [19] on NVIDIA GPUs (mention of commercial products does not constitute FDA endorsement; see Disclaimer). The resulting synthetic images are not paired to individual patient images and are designed to match the patient dataset at the distribution level. No demographic information or metadata was incorporated into the generation process. A total of 5,000 samples from CSAW-syn were used in this work.

The second synthetic dataset, VMLO-syn [20], was generated using StyleGAN2 trained on the MLO view subset of the VinDr-Mammo mammography patient dataset [21]. This subset contains approximately 10,000 images and was used exclusively, resulting in a generative model capable of producing MLO view mammograms. Training was conducted on NVIDIA A100 GPUs (mention of commercial products does not constitute FDA endorsement; see Disclaimer). As with CSAW-syn, the synthetic images produced by VMLO-syn are not paired to patient images and match the training data only at the distribution level. A total of 5,000 samples from VMLO-syn were used in this work.

### Head CT

Similar to digital mammography and WSI datasets, we also used three Head CT datasets in this study: one real-world patient dataset and two synthetic datasets.

The patient reference data for the head CT analysis consists of clinical head CT angiography (CTA) cases from the 2025 RSNA Intracranial Aneurysm Detection Challenge [22] annotated by expert neuroradiologists for aneurysm presence and 13 anatomical locations. A total of 1800 Head CTA volumes were acquired primarily from GE and Siemens as well as Philips, Toshiba, and Canon (mention of commercial products does not constitute FDA endorsement; see Disclaimer). Slice thickness ranges from 0.468 mm to 10.2 mm with an average of 1.23 mm, whereas average tube voltage is 120 kVp with a range between 70 and 150 kVp. Raw DICOM image intensities were converted to HU using the standard linear relationship defined by the dataset's rescale slope and intercept attributes. To obtain the segmentation masks for skull and brain volume (white and gray matter), we directly used TotalSegmentator [23] without any finetuning or retraining. To segment the Circle of Willis (CoW) segments in the artery vessels, we used the first-place solution [24] in the 2025 RSNA Intracranial Aneurysm Detection Challenge, which includes a vessel segmentation model trained specifically using the challenge dataset.

The synthetic head CTA dataset consists of 1,000 unique cases with the presence of an iodine-based contrast agent to make the arteries appear bright and highly visible. The process begins with the Multimodal Imaging-Based Detailed Anatomical (MIDA) phantom [25]. A knowledge-based approach is employed to model different anatomical variations, specifically implementing Circle of Willis variations to reflect real-world population-level variety [24, 26]. These customized digital phantoms serve as inputs for the XCIST simulation pipeline [27], which models physics and scanner configurations to produce the final synthetic CTA scan images.

Similar to the synthetic CTA dataset, the synthetic non-contrast head CT (NCCT) dataset consists of 1,000 unique cases derived from the same diversified MIDA digital phantoms. By turning off the simulated iodine contrast, this synthetic NCCT dataset is designed to have a fundamentally different attenuation profile and vessel texture feature set compared to the CTA data, as the unenhanced blood blends naturally with the surrounding soft tissues rather than appearing bright and highly defined. These non-contrast phantoms are then processed through the identical XCIST simulation pipeline to generate the final synthetic NCCT scan images. As with the WSI dataset, both CTA and NCCT synthetic data were generated for use in this study.

## RESULTS

### Feature Analysis and Metric Generation

Figure 3 illustrates the key steps in the ScoreCard pipeline using examples from 3 modalities: WSI, Mammography, and Head CT. Figure 3(a) presents the assessment of fidelity of synthetic histopathology image datasets generated by diffusion-based generative models (DDPMs) against a patient reference dataset, specifically the MoNuSeg dataset of patient pathology images. The pipeline flows from left to right across three main stages: Images & Feature Extraction, Feature Space Analysis, and Quantitative Comparison. Synthetic Dataset #1 (Coarse) was generated by a coarsely trained DDPM (trained broadly on real data without fine-tuning), while Synthetic Dataset #2 (Fine-tuned) was produced by a fine-tuned DDPM (further adapted to the real data distribution). Both datasets undergo knowledge-based feature extraction using the same computational pipeline applied to the patient data, ensuring methodological consistency towards a fair comparison. The extracted features are then projected into a two-dimensional PCA feature space, where convex hulls are drawn around the real and synthetic data point clouds to visualize their spatial overlap. In the case of the Coarse dataset, the synthetic convex hull is noticeably offset and poorly overlapping with the real hull, while the fine-tuned dataset shows substantially better alignment and overlap. These visual observations are quantified in the rightmost panel through metrics assessed along two complementary dimensions: Congruence, measured via three similarity scores, Jensen-Shannon Distance, Cosine Similarity, and Earth Mover's Distance, and Coverage, expressed as the Intersection over Union (IoU) of the convex hulls. The Coarse dataset returned lower values across all metrics (Cosine: 0.311, JSD: 0.555, EMD: 0.274, IoU: 0.231), indicating limited distributional alignment with the real data. By contrast, the fine-tuned dataset demonstrates markedly superior metric results (Cosine: 0.803, JSD: 0.132, EMD: 0.135, IoU: 0.650), suggesting that fine-tuning the generative model substantially improves the realism and distributional alignment of the synthetic images.

Figure 3(b) presents the assessment of fidelity of synthetic mammography image datasets generated by two distinct generative models against a patient reference dataset, specifically the EMBED (MLO) dataset of U.S. patient mammography images. Synthetic Dataset #1 (VMLO-Syn), generated using a StyleGAN2-based model trained on Vietnamese mammography data, while Synthetic Dataset #2 (Sinkove) was produced by the Sinkove pipeline using a diffusion model trained on Swedish data (mention of commercial products does not constitute FDA endorsement; see Disclaimer). All datasets undergo feature extraction using the same foundation model, DINOv2, ensuring methodological consistency and enabling a fair comparison. In the case of VMLO-Syn, the synthetic point cloud seen in the PCA plot is noticeably concentrated and poorly overlapping with the broader real hull, while Sinkove shows better distributional alignment and overlap with the real data distribution. The Congruence and Coverage metrics reflect the visual observations. VMLO-Syn showed intermediate metric values across all measures (Cosine: 0.784, JSD: 0.171, EMD: 0.082, IoU: 0.4103), with lower distributional alignment relative to Sinkove. By contrast, Sinkove demonstrates markedly higher metric values (Cosine: 0.902, JSD: 0.073, EMD: 0.049, IoU: 0.7582), suggesting that the Sinkove diffusion model, despite being trained on a geographically distinct Swedish population, more closely captures the distributional characteristics of the U.S. reference dataset than the StyleGAN2-based model trained on Vietnamese data. Figure 3(c) presents the fidelity assessment of two synthetic head CT datasets against the RSNA CTA patient data as the reference dataset. Both synthetic datasets undergo knowledge-based feature extraction applied consistently across patient and synthetic images, with results projected into a two-dimensional PCA feature space for visual and quantitative comparison. The mean intensity feature is used for metric computation.

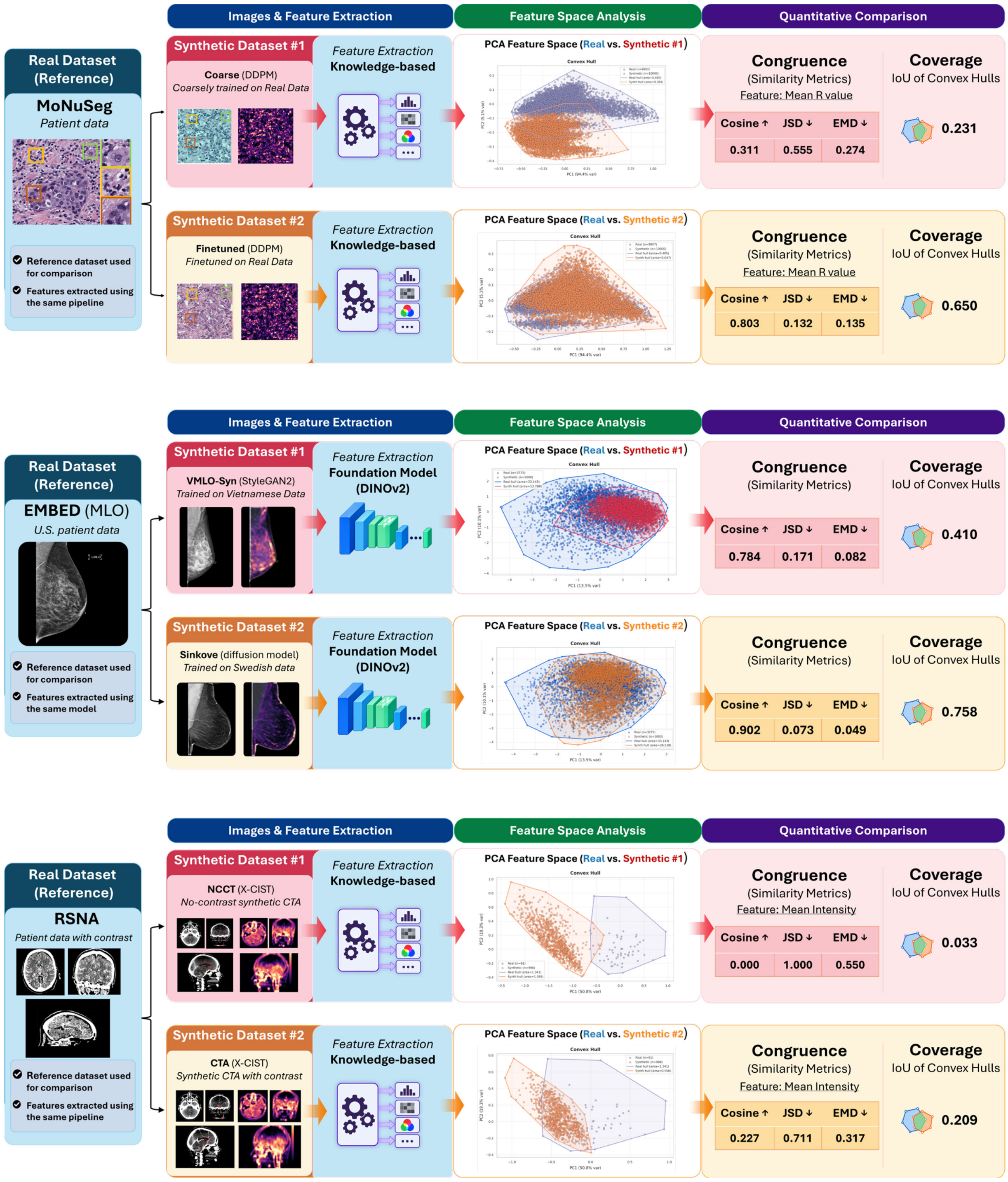

Figure 3: **End-to-end synthetic data evaluation pipeline across three medical imaging modalities.** For each modality, whole slide imaging (MoNuSeg), mammography (EMBED MLO), and brain CTA (RSNA), a real reference dataset is compared against two synthetic datasets. Features are extracted using either knowledge-based methods or foundation models, then projected into a PCA feature space to visualize distributional overlap between real (blue) and synthetic (orange) data. Quantitative comparison is performed using three similarity metrics, Cosine similarity, Jensen-Shannon Divergence (JSD), and Earth Mover's Distance (EMD), alongside an IoU-based Coverage score derived from convex hull overlap.

Synthetic Dataset #1 (NCCT) and Synthetic Dataset #2 (CTA) both showed low alignment with the real data distribution. In the PCA feature space, the synthetic data points form a tightly clustered, compact region in both cases, while the real data convex hull spans a substantially larger and broader area. This indicates that the synthetic datasets generated by XCIST capture only a narrow slice of the intensity variation present in the real reference images, failing to reproduce the full distributional spread of the RSNA dataset regardless of whether contrast is included in the generation process.

This is reflected in the quantitative metrics as well. NCCT shows near zero Congruence with the real data across all metrics representing essentially no distributional similarity to the RSNA dataset. CTA performs only marginally better in terms of Congruence (Cosine: 0.032, JSD: 0.956, EMD: 0.346), with negligible improvement despite the inclusion of contrast in the synthetic generation process. Both datasets showed low IoU-based Coverage values, consistent with limited distributional spread relative to the patient reference data.

Figure 4 shows a snapshot of the structured ScoreCard report generated for the WSI demonstration. The Dataset Overview summarizes information such as dataset name, modality, and number of samples, followed by feature-space visualizations, feature-level Congruence results reported with relative misalignment between the synthetic and patient data, and example images. Comprehension is represented using user-provided set of criteria. Coverage is visualized using overlapping convex hulls of the synthetic and patient distributions. The Constraint assessment identified relatively low but non-zero violation rates across the evaluated image features. Most synthetic feature values remained within the ranges observed in the patient reference data, while individual features exhibited violation rates of approximately 1–5%, most notably for appearance features such as mean intensity and RGB values. These results suggest that the synthetic images generally preserved the evaluated feature-level constraints, but retained a small number of samples with characteristics outside the patient-reference range. The Completeness assessment revealed a substantial difference in metadata availability between the synthetic and patient datasets. Approximately 44% of metadata values were missing in the synthetic dataset compared with approximately 12% in the patient dataset. Expected-field completeness was also lower for the synthetic data, at approximately 56%, compared with approximately 88% for the patient data. Thus, despite relatively strong image-level distributional alignment, the synthetic dataset contained substantially less complete accompanying information. The Consistency assessment showed substantial overlap in the distributions of normalized subgroup means for age at diagnosis between the patient and synthetic datasets. The median normalized subgroup mean was approximately 0.34 for the patient data and 0.28 for the synthetic data, with interquartile ranges of approximately 0.18–0.55 and 0.14–0.64, respectively. These results indicate a slightly lower central tendency and greater variability across subgroup means in the synthetic data. Additional plots for feature analysis, such as 2D PCA using all extracted features, can also be produced. In this demonstration, the combined outputs indicate generally strong alignment and feature coverage while also identifying smaller deviations in cell morphology and metadata uniformity. These observations illustrate the value of examining multiple quality dimensions: a dataset can show distributional similarity while retaining discrepancies that merit further review. For sample ScoreCard outputs for mammography and CTA, please refer to the supplementary materials.

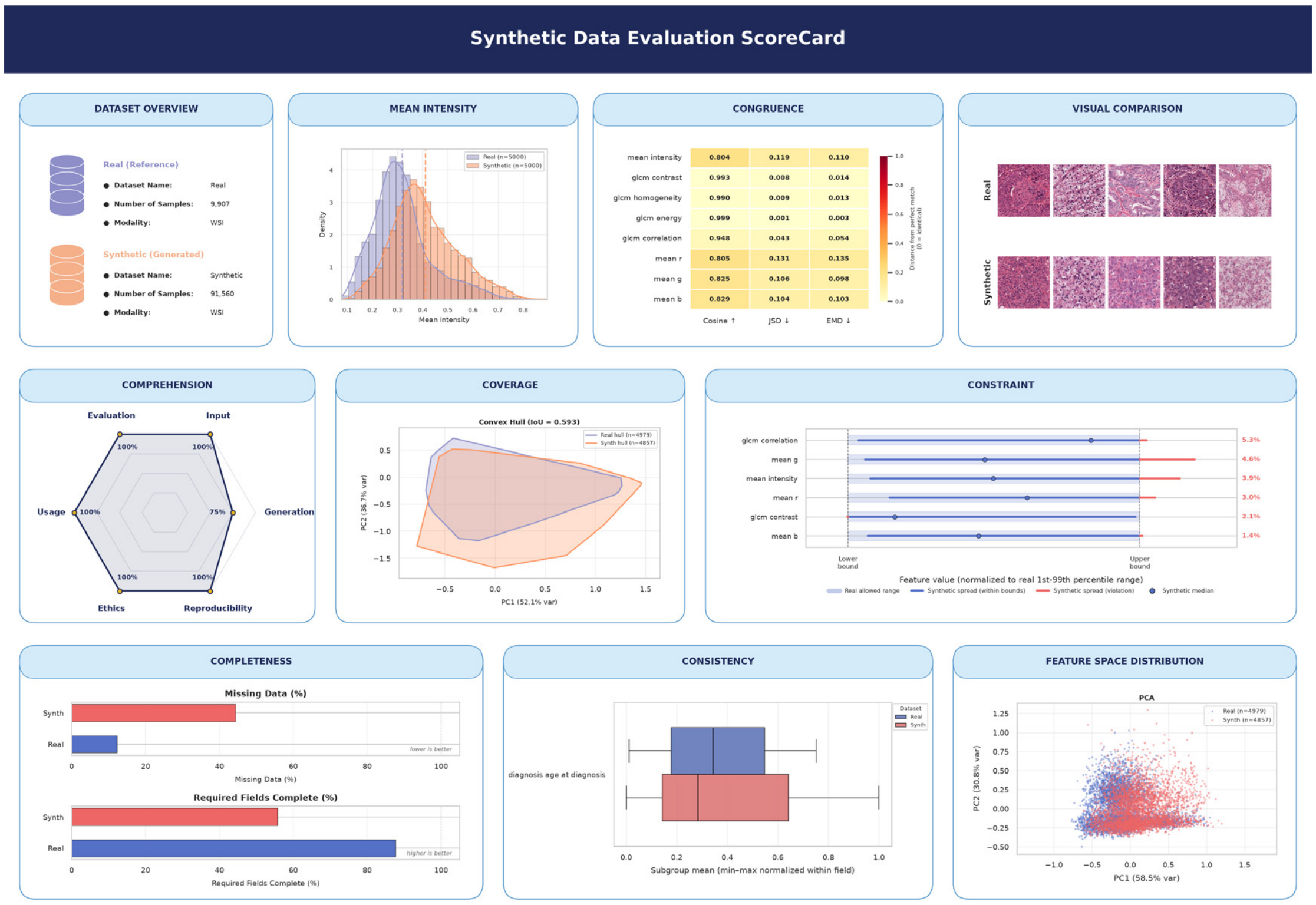

Figure 4: **Example ScoreCard output for a synthetic WSI dataset**.

## DISCUSSION

The ScoreCard demonstration of the 7Cs approach highlights an important distinction between distributional similarity and broader dataset quality. Congruence and Coverage describe how synthetic and patient data relate in the selected feature space, but these dimensions do not independently establish that synthetic images satisfy anatomical constraints, contain the metadata needed for the intended use, behave consistently across relevant subgroups, or are adequately documented. Completeness assesses whether the synthetic dataset contains the information needed for its intended use, including expected metadata fields, while Consistency examines whether these characteristics are maintained across relevant demographic or clinical subgroups. Constraint provides an additional check on plausibility by evaluating whether synthetic samples satisfy known anatomical, biological, geometric, or other application-specific characteristics. Comprehension complements these quantitative assessments by evaluating six aspects of documentation quality. Input describes the source data used for generation, and Generation describes the synthetic-data generation process. Reproducibility assesses whether the documentation includes the information needed to reproduce that process, while Ethics addresses relevant ethical considerations. Usage describes the intended applications and limitations of the synthetic data, and Features describes the characteristics represented in the synthetic dataset. Together, these dimensions extend the evaluation beyond distributional similarity by characterizing the usability, subgroup robustness, and interpretability of the synthetic data. The modular feature-extraction design is also important for interpretation. ScoreCard metrics operate on the representations supplied to them; consequently, the clinical relevance of an evaluation depends in part on whether the selected features capture properties that matter for the given imaging modality and intended use. The results from the three modalities in this paper highlight the usefulness of different feature categories, such as color, intensity, and foundation model for the assessment of synthetic data from various imaging domains. The implementation permits other feature representations, but this manuscript does not establish that any one feature space is sufficient across modalities or clinical tasks. Modality- and task-specific feature selection therefore remains an important consideration when applying the toolkit. Similarly, universal acceptance thresholds for ScoreCard metrics may not be appropriate because the interpretation of a given metric depends on the imaging modality, selected feature representation, intended use, and the clinical relevance of the characteristics being measured. Rather than defining fixed pass/fail criteria, the metrics are intended to provide quantitative evidence that can be interpreted in the context of the specific application and alongside expert judgment. By characterizing synthetic data quality both overall and across relevant subgroups, ScoreCard may help users identify dataset properties associated with differences in downstream task results, including subgroup-specific performance variation.

There are several limitations to this work. First, the end-to-end demonstrations use a single patient dataset for each modality which may lack sufficient variability due to either limited number of samples in the case of WSI and CTA or geographic limitations in case of mammography. Second, Compliance is intentionally not quantified by the toolkit, and Comprehension is a set of criteria rather than a quantitative image-derived measure. Third, the current results characterize dataset properties relative to the selected patient reference and feature representations; they do not establish clinical validity or suitability for a particular intended use. Finally, analyses relating ScoreCard metrics to downstream task performance are ongoing. Such analyses could clarify which quality dimensions and metrics are most informative for specific model development and evaluation scenarios. These limitations also define directions for continued development. Evaluation across additional modalities and synthetic data generation methods can test the portability of the pipeline, while downstream-task analyses can clarify how individual ScoreCard dimensions relate to model performance. Because the toolkit stores intermediate features and metric outputs and produces a structured report, it can also support reproducible comparisons as additional feature extractors and dimension-specific measures are incorporated. The present analysis does not establish that the evaluated synthetic data are suitable for a specific downstream clinical task.

## CONCLUSIONS

The ScoreCard Toolkit provides a practical and extensible implementation for evaluating and reporting SMD quality across the six dimensions of Congruence, Coverage, Completeness, Consistency, Constraint, Compliance, and Comprehension. By retaining multiple dimension-specific measures, the toolkit extends evaluation beyond fidelity-oriented comparisons and enables users to examine distributional alignment, coverage, constraint adherence, metadata completeness and consistency, and documentation-related information within a common reporting workflow. In our experiments across three medical imaging modalities, the ScoreCard output uncovered properties of synthetic datasets such as color bias in WSI, lack of distributional variety in Mammo potentially due to regional differences in the training data for the synthetic data generation pipeline, and low Coverage score in Phantom-based head CT datasets. Ongoing work includes evaluation on additional imaging modalities, completion of analyses relating ScoreCard metrics to downstream task performance, further exploration of subgroup analysis, and continued development of the open-source Python implementation.

## ACKNOWLEDGEMENT

The findings and conclusions in this article are those of the author(s) and do not necessarily represent the official position of the U.S. Food and Drug Administration (FDA), the Department of Health and Human Services (HHS), or the U.S. Government. Mention of trade names, commercial products or tools, or organizations does not imply endorsement by the U.S. Government.

## AUTHORS CONTRIBUTIONS

Seyed Kahaki: Conceptualized and designed the study, drafted the original manuscript, developed the code, data preparation, and performed analyses. Alexander Webber: Reviewed and revised the manuscript, developed the code, data preparation, trained and tested models, and conducted formal analyses. Tahsin Rahman: Reviewed and revised the manuscript, helped in developing the code, data preparation for digital mammography, and conducted formal analyses. Elim Thompson: Reviewed and revised the manuscript, provided synthetic head computer tomography data. Jana Delfino: Supervised the project and reviewed and revised the manuscript. Aldo Badano: Supervised the project and reviewed and revised the manuscript. Seyed Kahaki, Alexander Webber, and Tahsin Rahman contributed equally.

## Supplement 1

Table S1.1. Describes the features available for extraction from each sample in each of the three modalities explored in this work.

| TYPE | | FEATURE | WSI | MAMMOGRAPHY | CTA |
|---|---|---|---|---|---|
| GENERAL FEATURES | | Mean Intensity | ✓ | ✓ | ✓ |
| | | Mean RGB | ✓ | | |
| | | GLCM Contrast | ✓ | ✓ | ✓ |
| | | GLCM Homogeneity | ✓ | ✓ | ✓ |
| | | GLCM Energy | ✓ | ✓ | ✓ |
| | | GLCM Correlation | ✓ | ✓ | ✓ |
| DOMAIN SPECIFIC | Nuclei Topology | Nearest Neighbor | ✓ | | |
| | | K-Nearest Neighbors | ✓ | | |
| | | Radius Neighbors | ✓ | | |
| | | Voronoi Neighbors | ✓ | | |
| | Nuclei Shape | Convexity | ✓ | | |
| | | Roundness | ✓ | | |
| | | Solidity | ✓ | | |
| | | Eccentricity | ✓ | | |
| | | Skeleton Length | ✓ | | |
| | | Skeleton Complexity | ✓ | | |
| | Mammography Shape | Breast Contour | | ✓ | |
| | Vessel Geometry | Number of segments | | | ✓ |
| | | Total length | | | ✓ |
| | | Tortuosity | | | ✓ |
| | | Radius | | | ✓ |
| | | Curvature | | | ✓ |
| | | Torsion | | | ✓ |
| | | Tapering | | | ✓ |
| | | Number of bends | | | ✓ |
| | | DINOv2 | | ✓ | |
| | | UNI-2h | ✓ | | |
| | | Mammo-CLIP | | ✓ | |
| | | NYUMedML | | | ✓ |

Table S1.2. Description of the ScoreCard features.

| FEATURE | DESCRIPTION |
|---|---|
| **Mean Intensity** | Average image intensity within a given region |
| **Mean RGB** | Average image RGB colors |
| **GLCM Contrast** | Measures local intensity variation and differences between neighboring pixels |
| **GLCM Homogeneity** | Measures the similarity and uniformity of neighboring pixel intensities |
| **GLCM Energy** | Measures texture uniformity based on the distribution of intensity pairs |
| **GLCM Correlation** | Measures the linear relationship between neighboring pixel intensities |
| **Nearest Neighbor** | Distance to the closest neighboring nuclei |
| **K-Nearest Neighbors** | Distances to the K closest neighboring nuclei |
| **Radius Neighbors** | Number of neighboring nuclei within a defined radius |
| **Voronoi Neighbors** | Number of neighboring nuclei based on Voronoi adjacency |
| **Convexity** | Measures how closely the nuclei shape follows its convex hull |
| **Roundness** | Measures how closely the nuclei shape resembles a circle |
| **Solidity** | Ratio of nuclei area to its convex hull area |
| **Eccentricity** | Measures nuclei shape elongation |
| **Skeleton Length** | Length of skeleton defining nuclei |
| **Skeleton Complexity** | Measures branching and structural complexity of the nuclei skeleton |
| **Breast Contour** | Measures distribution of angle gradients along breast boundary |
| **Number of Segments** | Total number of vessel segments within the blood artery mask |
| **Total Length** | Total length of all vessel segments |
| **Tortuosity** | Arc length / chord length |
| **Radius** | Average, min, and max vessel radius |
| **Curvature** | Measurement of sharpness in bends |
| **Torsion** | Measurement of twisting out of a flat plane |
| **Tapering** | Measurement of vessel narrowing or widening |
| **Number of Bends** | Number of bends, calculated by summing CurveIDs |
| **DINOv2** | Deep image features extracted using the DINOv2 foundation model |
| **UNI-2h** | Deep image features extracted using the UNI-2h foundation model |
| **Mammo-CLIP** | Mammography-specific image features extracted using Mammo-CLIP foundation model |
| **NYUMEDML** | CT Brain-specific image features extracted using NYUMedML foundation model |

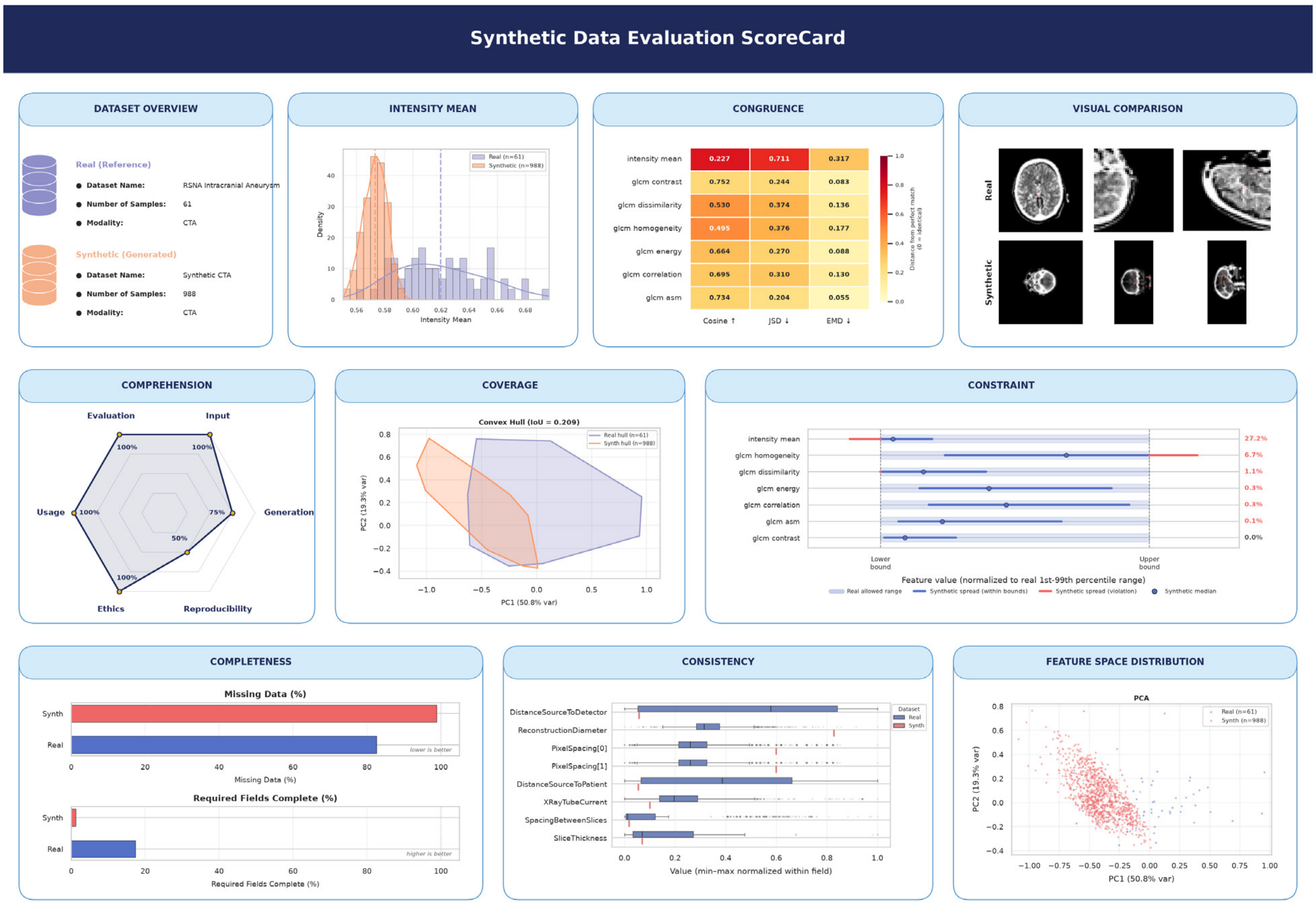

Figure S1.1 ScoreCard output for synthetic CTA.

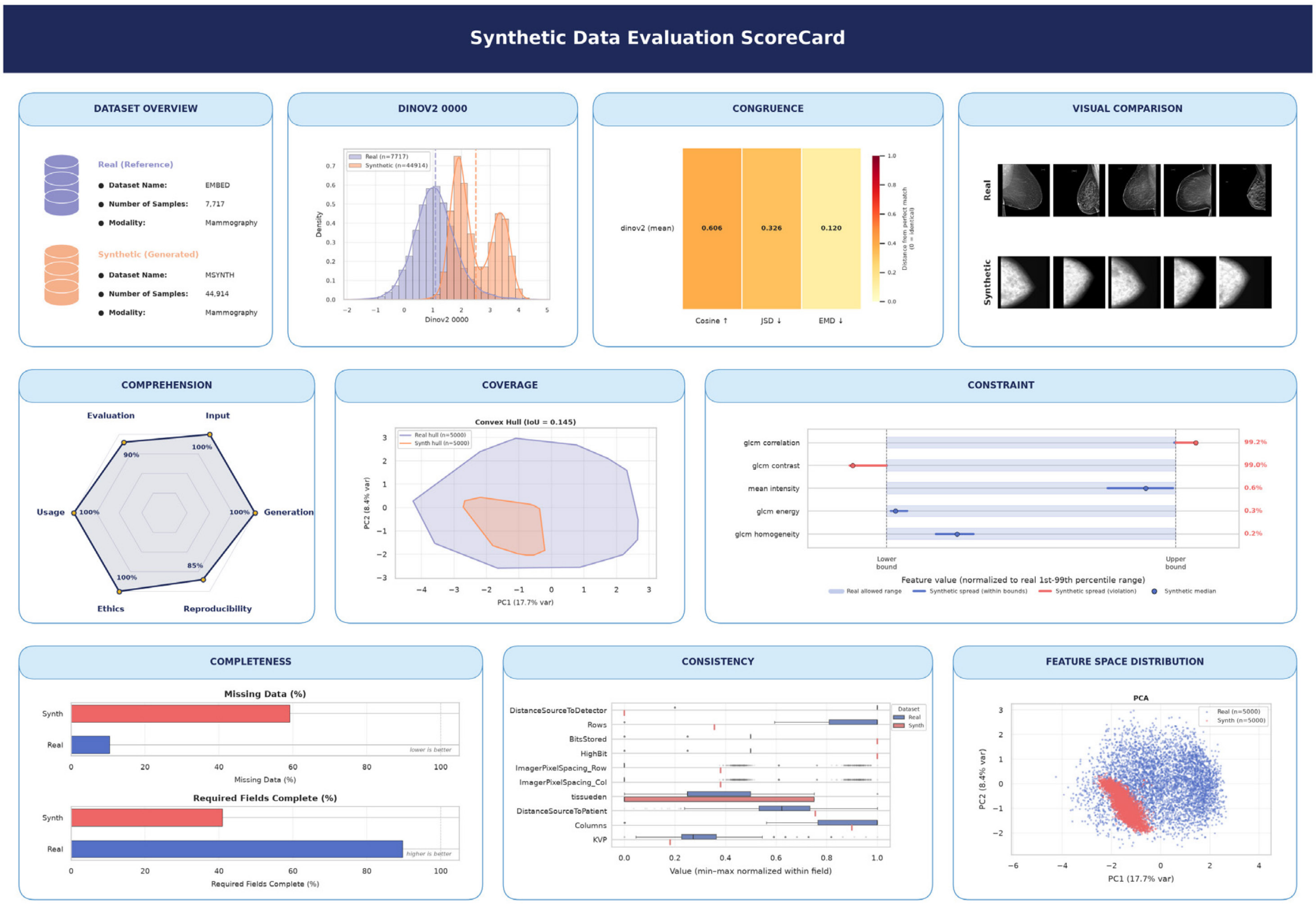

Figure S1.2 ScoreCard output for synthetic mammography dataset using DinoV2 feature extractors.